\documentclass[a4paper]{jpconf}
\usepackage{graphicx,cite}

\input{paperdef}

\begin{document}
\title{Central Exclusive Diffractive MSSM Higgs-Boson\\
Production at the LHC}

\author{S.~Heinemeyer$^1$,  V.A.~Khoze$^{2, 3}$, M.G.~Ryskin$^{2, 3}$, 
W.J.~Stirling$^{2, 4}$, M.~Tasevsky$^5$ and G.~Weiglein$^2$}

\address{
$^1$Instituto de Fisica de Cantabria (CSIC-UC), Santander, Spain\\
$^2$IPPP, Department of Physics, Durham University,
Durham DH1 3LE, U.K.\\
$^3$Petersburg Nuclear Physics Institute, Gatchina,
St.~Petersburg, 188300, Russia\\
$^4$Department of Mathematical Sciences,
Durham University, DH1 3LE, U.K.\\
$^5$Institute of Physics of the ASCR, v.~v.~i., Na Slovance 2, CZ-18221,
   Czech Republic
}

\ead{Georg.Weiglein@durham.ac.uk}

\begin{abstract}
The prospects for central exclusive diffractive (CED) production of 
MSSM Higgs bosons at the LHC are reviewed. It is shown that the CED
channels, making use of forward proton detectors at the LHC
installed at 220~m and 420~m distance around ATLAS and / or CMS,
can provide important information on the Higgs sector of the MSSM.
In particular, CED production of the
neutral $\cp$-even Higgs bosons $h$ and $H$ and their decays into
bottom quarks has the potential to probe interesting
regions of the $\MA$--$\tb$ parameter plane of the MSSM and may give access
to the bottom Yukawa
couplings of the Higgs bosons up to masses of $\MH \lsim 250 \gev$.
\end{abstract}

%%%%%%%%%%%%%%%%%%%%%%%%%%%%%%%%%%%%%%%%%%%%%%%%%%%%%%%%%%%%%%%%%%%%%%%%%%%%%%
%%%%%%%%%%%%%%%%%%%%%%%%%%%%%%%%%%%%%%%%%%%%%%%%%%%%%%%%%%%%%%%%%%%%%%%%%%%%%%

\newcommand{\sixoo}{60 \ifb}
\newcommand{\sixooeff}{60 \ifb\,eff$\times2$}
\newcommand{\sixooo}{600 \ifb}
\newcommand{\sixoooeff}{600 \ifb\,eff$\times2$}

\section{Introduction}

Searches for Higgs bosons and the investigation of their properties are
among the main goals of the LHC~\cite{atlastdr,atlasrev,cmstdr}. 
While the Standard Model (SM) comprises
only one Higgs boson, many models of new physics require an extended
Higgs sector. The most popular extension of the SM is the Minimal
Supersymmetric Standard Model (MSSM), whose Higgs sector consists of two
doublets, leading to five physical states. At lowest order the Higgs
sector of the MSSM is $\cp$-conserving, containing two $\cp$-even Higgs
bosons, $h$ and $H$, a  $\cp$-odd Higgs boson, $A$, and the charged
Higgs bosons $H^\pm$. The MSSM Higgs sector can be specified at lowest
order in terms of the gauge couplings, the ratio of the two Higgs vacuum
expectation values, $\tb \equiv v_2/v_1$, and the mass of the $\cp$-odd
Higgs boson, $\MA$.
Higgs physics in the MSSM is affected by large higher-order
corrections (see for example\ \citere{reviews} for recent
reviews), which
have to be taken into account for reliable phenomenological predictions.
Revealing that a detected new state is indeed a Higgs boson and
distinguishing the Higgs boson(s) of the SM or the MSSM from the states
of other theories will be non-trivial. This goal will require
a comprehensive programme of precision Higgs measurements. In
particular, it will be of utmost importance to determine the spin and
$\cp$ properties of a new state and to measure precisely its mass, width
and couplings.

The ``conventional'' LHC production channels, gluon fusion, weak boson
fusion and associated production with
heavy quarks or vector bosons, could be complemented by ``central
exclusive diffractive'' (CED) Higgs-boson production making use of
of forward proton taggers (Roman Pot (RP) detectors) installed at 
220~m and 420~m distance around
ATLAS and / or CMS~\cite{totemRP220,CMS-Totem}. In the exclusive processes
$pp\to p \oplus H \oplus p$, where the $\oplus$ signs are used to denote the 
presence of large rapidity gaps, there is no
hadronic activity between the outgoing protons and the decay products
of the central system. If the outgoing protons remain intact and scatter
through small
angles then, to a very good approximation, the primary active di-gluon
system obeys a $J_z=0$, $\cp$-even selection rule~\cite{KMRmm}. 
Here $J_z$ is the projection of the total
angular momentum along the proton beam axis. This selection rule
readily permits a clean determination of the quantum numbers of the
observed Higgs resonance which  will be dominantly produced in a
scalar state. Furthermore, because the process is exclusive, the energy
loss of the outgoing protons is directly related to the mass of the
central system, allowing a potentially excellent mass resolution,
irrespective of the decay mode of the produced particle. The CED
processes would allow  all the main
Higgs-boson decay modes, $b \bar b$,  $WW$  and $\tau\tau$, to be
observed in the same production channel. This could provide a unique
possibility to study the Higgs coupling to bottom quarks, which may be
difficult
to access in other search channels at the 
LHC~\cite{atlastdr,atlasrev,cmstdr} despite the fact that
$H \to b \bar b$ is by far the dominant decay mode for a light SM-like
Higgs boson.

Within the MSSM, CED Higgs-boson production can be even more important
than in the SM. The coupling of the lightest MSSM Higgs boson to bottom
quarks and $\tau$~leptons can be strongly enhanced for large values of
$\tb$ and relatively small values of $\MA$. On the other hand, for
larger values of $\MA$ the branching ratio of the heavy $\cp$-even Higgs
boson into bottom quarks, ${\rm BR}(H \to b \bar b)$, is much larger
than for a SM Higgs boson of the same mass. As a consequence, CED Higgs
boson production with decay to $b \bar b$ can be utilised in the MSSM up
to much higher Higgs-boson masses than in the SM case. 
We briefly review in the
following some of the results of \citere{higgsdiffract}, where the 
prospects for
CED production of MSSM Higgs bosons has been analysed in detail (see
also \citeres{KKMRext,otherMSSM,clp} for other studies in the MSSM).

%%%%%%%%%%%%%%%%%%%%%%%%%%%%%%%%%%%%%%%%%%%%%%%%%%%%%%%%%%%%%%%%%%%%%%%%%%%%%%%
%%%%%%%%%%%%%%%%%%%%%%%%%%%%%%%%%%%%%%%%%%%%%%%%%%%%%%%%%%%%%%%%%%%%%%%%%%%%%%%

\section{CED MSSM Higgs-boson production: cross sections, background processes
and experimental aspects}
\label{sec:backgrounds}

The Higgs signal and background cross sections can be written as a
function of the Higgs-boson mass with the help of simple approximate
formulae~\cite{KKMRcentr,KKMRext,higgsdiffract}.
The cross sections $\si^{\rm excl}$ for
CED production of $h, H$ can be obtained from
\begin{equation}
\si^{\rm excl} \, \mbox{BR}^{\rm MSSM} =3 \, {\rm fb} \left(\frac{136}{16+M}\right)^{3.3}
  \left(\frac{120}M\right)^3 
  \frac{\Ga(h/H \to gg)}{0.25\mev} \,\mbox{BR}^{\rm MSSM},
\label{eq1}
\end{equation}
where we calculate the gluonic partial width $\Ga(h/H\to gg)$ and the 
branching ratios for the various channels in the MSSM,
$\mbox{BR}^{\rm MSSM}$, using the
program \fh~\cite{feynhiggs}. 
The mass $M$ (in GeV) denotes either $\Mh$ or $\MH$.
The factor $(136/(16+M))^{3.3}$ accounts for the mass dependence of the
effective ``exclusive" $gg^{PP}$ luminosity, see 
\citeres{KKMRcentr,KKMRext}.
The normalisation is fixed  at $M=120 \gev$, where in accordance with 
\citere{KKMRext} we obtain $\si^{\rm excl} =3$~fb with the width  
$\Ga(\HSM \to gg)=0.25 \mev$.
In \citere{KKMRext} various uncertainties in the prediction of the
CED cross sections were discussed, leading to an estimate of an
uncertainty factor of $\sim 2.5$ (see also \citere{higgsdiffract} for a more
detailed discussion)%
\footnote{Additional uncertainties in the production cross sections of
up to $\sim 20\%$ could arise for large $\tb$ due to the imperfect
inclusion of NNLO QCD corrections.}%
.~Eq.~(\ref{eq1}) yields a total number of signal events of about~100 for a SM
Higgs boson with $\MHSM = 120 \gev$ with an integrated luminosity of 
60~fb$^{-1}$ if only the
forward detector acceptances are accounted for and no cuts and
efficiencies in the central detector are imposed (summing over the
different Higgs decay channels).

Within the accuracy of the existing calculations
\cite{KMRmm,DKMOR,krs2},
the overall background to the $0^+$ Higgs signal in the
$b \bar b$ mode can be approximated by
\BE
\frac{{\rm d}\si^B}{{\rm d} M} \approx 0.5 \, {\rm fb/GeV} \left[
0.92\left(\frac{120}{M}\right)^6 +
\frac{1}{2} \left(\frac{120}{M}\right)^8
                                  \right] .
\label{eq:backbb}
\EE
This expression summarises several types of background subprocess (see
\citeres{DKMOR,bh75,krs2} and the discussion in \citere{higgsdiffract}):
the prolific (LO)
$gg^{PP}\to gg$ subprocess can mimic $b\bar b$ production since one may
misidentify the outgoing gluons as $b$ and $\bar{b}$ jets;
an admixture of $|J_z|=2$ production, arising from non-forward going
protons, which contributes to the (quark-helicity conserving) 
LO $gg^{PP}\to b\bar b$
background;
since the $b$-quarks have non-zero mass there is a contribution to the
$J_z=0$ (quark-helicity non-conserving) cross section of order $\mb^2/E_T^2$;
there is the possibility of NLO $gg^{PP}\to b\bar b g$
(dominantly quark-helicity conserving) background contributions, which for hard
gluon radiation at large angles do not obey the selection rules;
another potential background source can arise from
the interaction of two soft Pomerons.

At high instantaneous luminosity, i.e.\ 
$10^{34} \, {\rm cm}^{-2} \, {\rm s}^{-1}$, the main experimental
challenge will be pile-up events that contain protons within the
acceptances of the RPs.
While single pile-up events would not survive the signal selection cuts
(see below), the
overlay of two single diffractive events with a hard-scale inclusive 
non-diffractive event in the central detector could mimic the signal.
The pile-up issue is currently under intense study within ATLAS
and CMS (for detailed discussions, see \citeres{CMS-Totem,clp}). Possible
leverages that could bring the pile-up problem under control are
fast timing detectors, precise vertex detectors, the fact that 
signal and pile-up events possess different track multiplicity
properties, and a matching of the whole 4-momentum of the central system
(measured in the central detector) to the corresponding values obtained
from the forward proton taggers.

At nominal LHC optics, forward proton taggers positioned at a distance
$\pm 420$~m from the interaction points of ATLAS and CMS will allow
coverage in the proton
fractional momentum loss $\xi$ in the range 0.002--0.02,
with an acceptance of around 30\% for a centrally produced system
with a mass around $120 \gev$.
A combination with the foreseen proton detectors at
$\pm 220$~m~\cite{totemRP220} would
significantly increase the physics reach of forward studies enlarging
the $\xi$ range up to 0.2.
This would be especially beneficial because of
the acceptance for higher mass states and improvements in the
triggering~\cite{CMS-Totem}. 
%The three possible RP configurations are denoted as ``420''
%(for a proton detection at 420~$+$~420~m), ``combined'' (for the
%detection at 220~$+$~420~m) and ``220'' (for the detection at
%220~$+$~220~m).

The main selection criteria for $h,H \to b \bar b$ are either two
$b$-tagged jets or two jets with at least one $b$-hadron decaying
into a muon.
Details on the corresponding selection cuts can be found
in \citeres{higgsdiffract,CMS-Totem}.
To retain the signal at the Level~1 trigger,
the following trigger conditions can be used:
(1)
Single-sided 220~m RP and at least two jets, each with
{$E_{\rm T}>40 \gev$}, measured in the central detector.
(2)
A jet with {$E_{\rm T}>40 \gev$} and at least one muon with
{$E_{\rm T}>3 \gev$}, both measured in the central detector.
(3)
At least two jets each with {$E_{\rm T}>90 \gev$} measured in the
central detector.
(4)
Leptonic triggers, requiring electrons or muons in the central
detector.

For the process $h,H \to b \bar b$, a combination of the triggers~1
and~2 allows
the retention of about 65\% of the signal events passing the relevant
cuts
at $M = 120 \gev$ and up to 100\% at
$M = 200 \gev$, while at masses well above $200 \gev$ the trigger~3
retains the whole signal sample selected by the cuts.

In the numerical analysis below we will consider four luminosity
scenarios, \sixoo, \sixooeff, \sixooo\ and \sixoooeff. Here 
\sixoo\ and \sixooo\ refer to running at low and high 
instantaneous luminosity, respectively, using conservative assumptions
for the signal rates and the experimental
sensitivities~\cite{higgsdiffract}. Improvements on the experimental
side and possibly higher signal rates could lead to scenarios where the
event rates are higher by a factor of 2, denoted as \sixooeff\ and
\sixoooeff.

%%%%%%%%%%%%%%%%%%%%%%%%%%%%%%%%%%%%%%%%%%%%%%%%%%%%%%%%%%%%%%%%%%%%%%%%%%%%%%%
%%%%%%%%%%%%%%%%%%%%%%%%%%%%%%%%%%%%%%%%%%%%%%%%%%%%%%%%%%%%%%%%%%%%%%%%%%%%%%%

\section{Discovery reach}
\label{sec:discovery}

\begin{figure}[htb!]
%\vspace{1em}
\BC
\includegraphics[width=.495\textwidth]{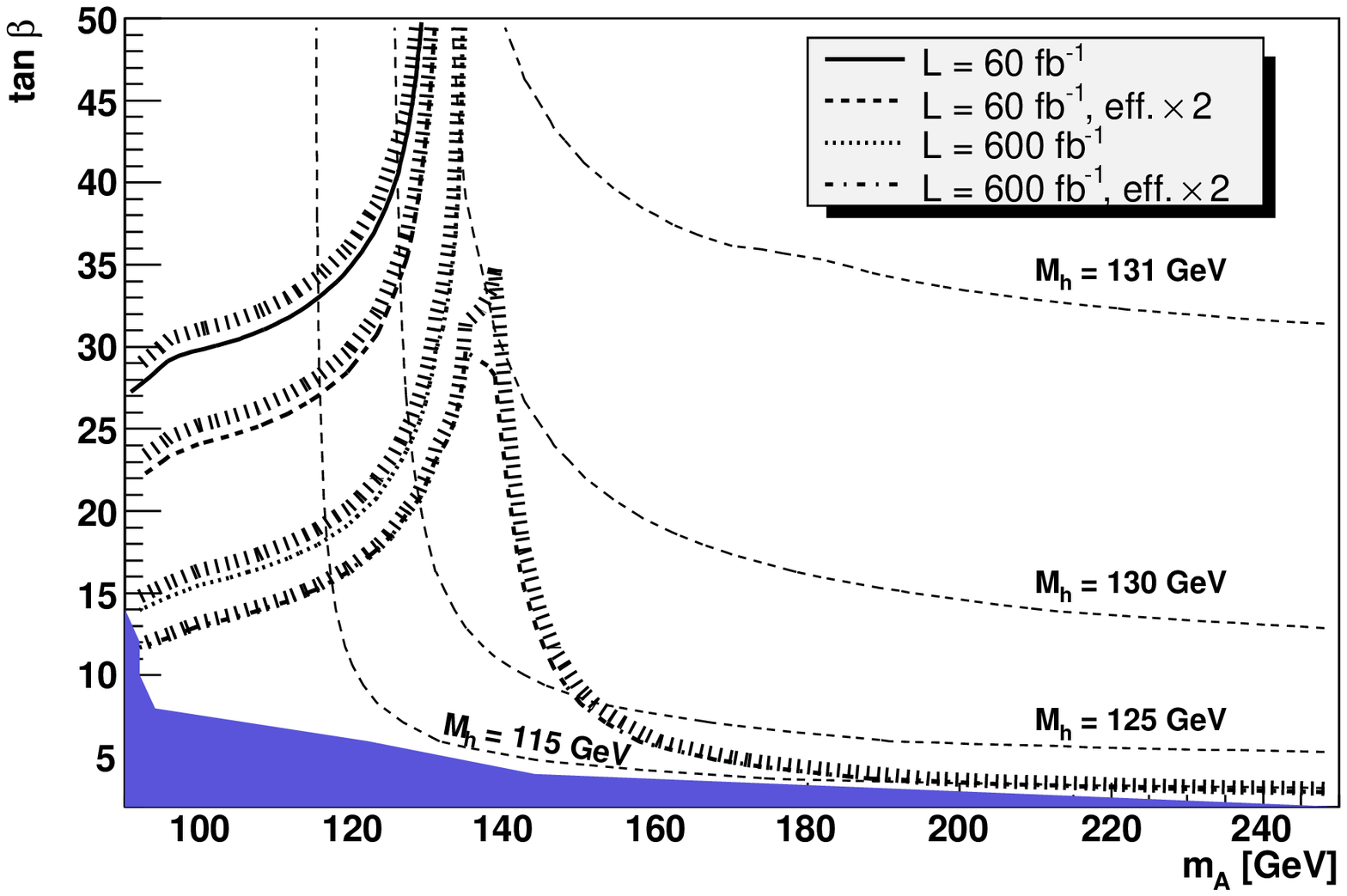}
\includegraphics[width=.495\textwidth]{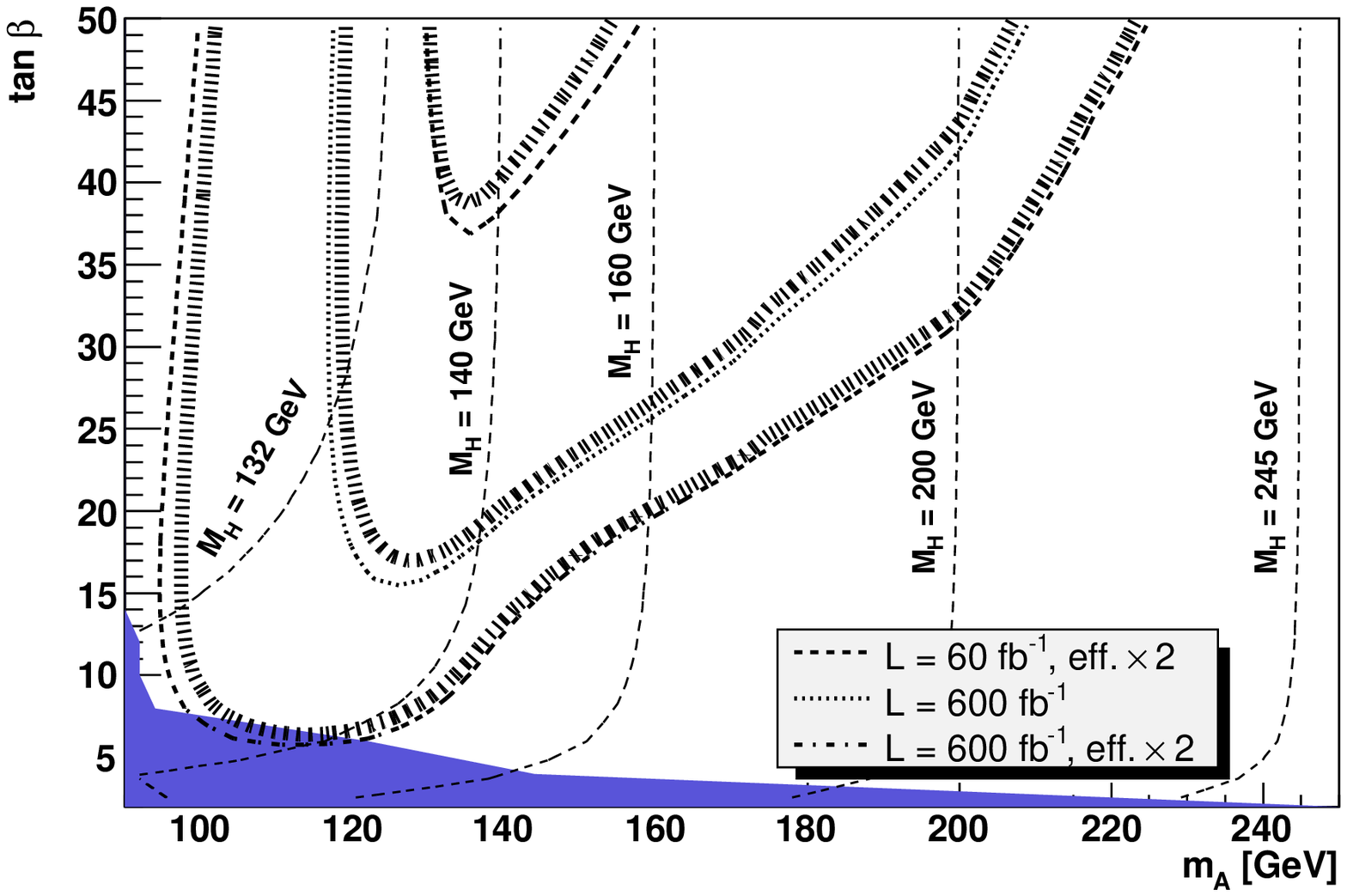}
\EC
\vspace{-1em}
\caption{
$3 \si$ statistical significance contours for the $h \to b \bar b$
channel (left plot) 
and
$5 \si$ discovery contours for the $H \to b \bar b$ channel (right
plot)
in CED production in the $\MA$--$\tb$ plane of the MSSM within the
$\Mhmax$ benchmark scenario with $\mu = +200 \gev$. The results are shown for
assumed effective
luminosities (see text, combining ATLAS and CMS) of \sixoo,
\sixooeff, \sixooo\ and \sixoooeff.
The values of $\Mh$ (left plot) and $\MH$ (right plot) are
indicated by contour lines. The dark shaded (blue)
region corresponds to the parameter region that
is excluded by the LEP Higgs
searches in the channel
$e^+e^- \to Z^* \to Z h, H$~\cite{LEPHiggs}.
No exclusion limits from the Higgs searches at the Tevatron are shown,
since at present the parameter region with
$\MA \gsim 100 \gev$ and $\tb \lsim 50$ is hardly affected by
the Tevatron exclusion bounds.
}
\label{fig:disc}
\vspace{-1em}
\end{figure}

We now discuss the prospects for CED production of the neutral
$\cp$-even MSSM Higgs bosons in the $\MA$--$\tb$ plane 
of the $\Mhmax$ benchmark scenario~\cite{benchmark}.
Since the lighter $\cp$-even Higgs boson of the MSSM is likely to be
detectable also in ``conventional'' Higgs search channels at the LHC
(see for example \citeres{atlastdr,cmstdr}), it may not be necessary to
require a statistical significance as high as $5 \si$
for the CED channel. Therefore in the left plot of 
\reffi{fig:disc}
we display contours of $3 \si$
statistical significances for the four luminosity
scenarios defined in \refse{sec:backgrounds}.
While the region of high $\tb$ and low $\MA$ can be probed also with 
lower integrated luminosity, in the \sixoooeff\ scenario the coverage
extends over nearly the whole $\MA$--$\tb$ plane, with the exception of
a small parameter region around $\MA \approx 140\gev$. The coverage
therefore includes the case of a light SM-like Higgs, which corresponds
to the region of large $\MA$ in the plot. As a consequence, if
the CED channel can be utilised at high instantaneous
luminosity (which requires in particular that pile-up background
is brought under control, see the discussion above) this channel can 
contribute very important
information on the Higgs sector of the MSSM. Besides giving access to
the bottom Yukawa coupling, which is a crucial input for determining all
other Higgs-boson couplings~\cite{HcoupLHCSM}, observation of a Higgs
boson in CED production with subsequent decay into bottom quarks would
provide information on the $\cp$ quantum numbers of the new state, yield
an (additional)
precise mass measurement, and may even allow a direct measurement of the
Higgs-boson width.

%%%%%%%%%%%%%%%%%% F I G U R E %%%%%%%%%%%%%%%%%%%%%%%%%%%%%%%%%%%%%%%%%%%%%%
%%%%%%%%%%%%%%%%%% F I G U R E %%%%%%%%%%%%%%%%%%%%%%%%%%%%%%%%%%%%%%%%%%%%%%

The properties of the heavier $\cp$-even Higgs boson
of the MSSM differ very significantly from the ones of a SM Higgs in
the region where $\MH, \MA \gsim 150 \gev$. While for a SM Higgs the
$\mbox{BR}(H \to b \bar b)$ is strongly suppressed in this mass region, the
decay into bottom quarks is the dominant decay mode for the heavier
$\cp$-even MSSM Higgs boson (as long as no decays into supersymmetric
particles or lighter Higgs bosons are open). 
The discovery reach in the
``conventional'' search channels at the LHC, in particular
$b\bar b H/A, H/A \to \tau^+\tau^-$, covers the parameter region of high
$\tb$ and not too large $\MA$~\cite{atlastdr,atlasrev,cmstdr},
while a ``wedge region''~\cite{atlastdr,cmstdr,cmsHiggs} remains
where the heavy MSSM Higgs bosons escape detection at the LHC.

The right plot of \reffi{fig:disc} shows the 
$5 \si$ discovery contours for the $H \to b \bar b$ channel in the 
$\Mhmax$ scenario. 
In the ``\sixoooeff'' scenario the discovery reach for the
heavier $\cp$-even Higgs boson goes beyond $\MH \approx 200 \gev$ in
the large $\tb$ region at the $5 \si$ level (at the $3 \si$ level, see
\citere{higgsdiffract},
the coverage extends to about $\MH = 250 \gev$ for $\tb \approx 50$).
Thus, CED production of the heavier
$\cp$-even Higgs boson of the MSSM with subsequent decay into bottom
quarks provides a unique opportunity for accessing its bottom Yukawa
coupling in a mass range where for a SM Higgs boson the decay rate into
bottom quarks would be negligibly small.
It is interesting to note that
in the ``\sixoooeff'' scenario the ($5 \si$ level)
discovery of a heavy $\cp$-even
Higgs boson with a mass of about $140 \gev$ will be possible for all
values of $\tb$. 

In conclusion,
the CED channels have an interesting physics potential for
probing the MSSM Higgs sector. Further experimental and theoretical
efforts in exploring this possibility are desirable.

\subsection*{Acknowledgements}

Work supported in part by the European Community's Marie-Curie Research
Training Network under contract MRTN-CT-2006-035505
`Tools and Precision Calculations for Physics Discoveries at Colliders'
(HEPTOOLS).

\vspace{1em}

%%%%%%%%%%%%%%%%%%%%%%%%%%%%%%%%%%%%%%%%%%%%%%%%%%%%%%%%%%%%%%%%%%%%%%%%%%%%%%
%%%%%%%%%%%%%%%%%%%%%%%%%%%%%%%%%%%%%%%%%%%%%%%%%%%%%%%%%%%%%%%%%%%%%%%%%%%%%%

% Non-BibTeX users please use

%%%%%%%%%%%%%%%%%%%%%%%%%%%%%%%%%%%%%%%%%%%%%%%%%%%%%%%%%%%%%%%%%%%%%%%%%%%%%%
%%%%%%%%%%%%%%%%%%%%%%%%%%%%%%%%%%%%%%%%%%%%%%%%%%%%%%%%%%%%%%%%%%%%%%%%%%%%%%

\end{document}